\documentclass[11pt,a4paper]{article}

\usepackage{epsfig}
\usepackage{amssymb}
\usepackage{amsmath}
\usepackage{ae}

\begin{document}

\title{Gravitational lensing by Einstein-Born-Infeld black holes}

\author{Ernesto F. Eiroa\thanks{e-mail: eiroa@iafe.uba.ar}\\
{\small Instituto de Astronom\'{\i}a y F\'{\i}sica del Espacio, C.C. 67, 
Suc. 28, 1428, Buenos Aires, Argentina}}

\maketitle
\date{}

\begin{abstract}
In this paper, charged black holes in general relativity coupled to 
Born--Infeld electrodynamics are studied as gravitational lenses. The 
positions and magnifications of the relativistic images are obtained 
using the strong deflection limit, and the results are compared with those 
corresponding to a Reissner--Nordstr\"{o}m black hole with the same 
mass and charge. As numerical examples, the model is applied to the 
supermassive Galactic center black hole and to a small size black hole 
situated in the Galactic halo.
\end{abstract}



\section{Introduction}

The study of gravitational deflection of light by a compact object with a 
photon sphere requires a full strong field treatment, instead of the weak 
field approximation \cite{schne} (i.e., only keeping the first non null term 
in the expansion of the deflection angle), commonly  used for ordinary stars 
and galaxies. Light rays passing close to the photon sphere will have large 
deflection angles, resulting in the formation of two infinite sets of faint 
relativistic images produced by photons that make complete turns (in both 
directions of rotation) around the black hole before reaching the observer, 
in addition to the primary and secondary weak field images. In the last few 
years, there has been a growing interest in strong field lensing situations. 
Virbhadra and Ellis \cite{virbha1} introduced a lens equation for 
asymptotically flat spacetimes and made a numerical analysis of lensing 
due to the black hole situated in the center of the Galaxy, using the
Schwarzschild metric. In another article \cite{virbha2}, they investigated 
gravitational lensing by naked singularities. Fritelli, Kling and Newman 
\cite{fritelli} found an exact lens equation without any reference to a 
background metric and compared their results with those of Virbhadra and 
Ellis. A logarithmic approximation was used by several authors \cite{prev} 
to obtain the deflection angle as a function of the impact parameter in 
Schwarzschild geometry, for light rays passing very close to the photon 
sphere, in the analysis of strong field situations. This asymptotic 
approximation is the starting point of an analytical method for strong field 
lensing, called the strong field limit by Bozza \textit{et al.} \cite {bozza1} 
or following a suggestion by Perlick \cite{perlick}, the strong deflection 
limit, which gives the lensing observables in a straightforward way 
\cite {bozza1}. Eiroa, Romero and Torres \cite{eiroto} extended this method to 
Reissner-Nordstr\"{o}m geometry, and Bozza \cite{bozza2} showed that it can be 
applied to any static spherically symmetric lens. It was subsequently used by 
Bhadra \cite{bhadra} to study a charged black hole lens of string theory, 
by Petters \cite{petters} to analyze the relativistic corrections to 
microlensing effects produced by the Galactic black hole, and by Bozza and 
Mancini \cite{bozman1} to study the time delay between different relativistic 
images. All the above mentioned works treated standard lensing situations, 
i.e. the lens is placed between the source and the observer. But when the 
observer is placed between the source and the black hole lens, or the source 
is situated between the lens and the observer, two infinite sequences of 
images with deflection angles closer to odd multiples of $\pi$ are formed, a 
situation called retrolensing. Holtz and Wheeler \cite{holtz} considered a 
black hole retrolens in the galactic bulge with the sun as source, and 
De Paolis \textit{et al.} \cite{depaolis1} analyzed the massive black hole at 
the Galactic center as retrolens with the bright star S2 as source; in both 
works, the Schwarzschild metric was used and only the two strongest images 
were taken into account. Eiroa and Torres \cite{eitor} studied the general 
case of a spherically symmetric retrolens, using the strong 
deflection limit to obtain the positions and magnifications of all images. 
Bozza and Mancini \cite{bozman2} extended the strong deflection limit 
to analyze standard lensing, retrolensing and intermediate situations under a 
unified formalism, and analyzed several stars in the neighborhood of the 
central Galactic black hole as possible sources. The study of lensing by 
rotating black holes is more complicated than by spherically symmetric 
ones. In recent years, some works \cite{bozza3,kl} considered spinning 
black hole lenses, in most cases restricting their treatment to equatorial or 
quasi-equatorial lensing scenarios. The complete extension of the 
strong deflection limit to Kerr geometry was not possible yet, but 
important advances in this direction were made by Bozza \textit{et al.} 
\cite{bozza3}. The relativistic images produced by spherically 
symmetric black holes lenses in the context of braneworld cosmologies were 
investigated by Eiroa \cite{eiroa} and Whisker \cite{whisker}. Other related 
topics about strong field lensing are treated in Refs. \cite{other}.\\

Born and Infeld \cite{borninf} proposed in 1934 a nonlinear theory of 
electrodynamics in order to avoid the infinite self energies for charged point 
particles arising in Maxwell theory. In 1935, Hoffmann \cite{hoffmann}  
joined general relativity with Born--Infeld electrodynamics to obtain a 
spherically symmetric solution representing the gravitational field of a 
charged object. These works were nearly forgotten for several decades, until 
the interest in non linear electrodynamics increased in the context of low 
energy string theory, in which Born--Infeld type actions appeared \cite{nle}. 
Gibbons and Rasheed showed \cite{gibbons} that Maxwell and Born--Infeld 
theories are singled out among all electromagnetic theories for having 
electric-magnetic duality invariance. Hoffmann solution failed to represent a 
suitable classical field model for the electron, instead it corresponds to 
that is now called a black hole. Spherically symmetric black holes in non 
linear electrodynamics coupled to Einstein gravity were studied in recent 
years by several authors \cite{gibbons,nlebh}. Pleba\~{n}ski \cite{pleb} 
found that in Born--Infeld electrodynamics the trajectories of photons in 
curved spacetimes are not null geodesics of the background metric. Instead, 
they follow null geodesics of an effective geometry determined by the 
nonlinearities of the electromagnetic field. Gutierrez \textit{et al.} 
\cite{gutpleb} and Novello \textit{et al.} \cite{novello} extended the concept 
of effective metrics for photons to any nonlinear electromagnetic theory.  
Bret\'{o}n \cite{breton} analyzed the geodesic structure of 
Einstein--Born--Infeld black holes for massive particles, photons and 
gravitons.\\

It is commonly thought that astrophysical black holes have no charge, because 
selective accretion of charge will tend to neutralize them if they are 
situated in a high density environment. But in recent years some mechanisms 
were proposed that could produce charged black holes. It was argued 
\cite{punsly} that charged rotating black holes can exist if they are 
surrounded by a co-rotating magnetosphere with equal and opposite charge. 
These black holes could survive for a long time if they are in a low density 
medium. A model in which the presence of a strong and high energy radiation 
field may induce an electric charge into an accreting black hole was presented 
by de Diego \textit{et al.} \cite{diego}. Ghezzi and Letelier \cite{ghele} 
made a numerical simulation of stellar core collapse resulting in the 
formation of Reissner--Nordstr\"{o}m spacetimes. Within braneworld 
cosmologies, Mosquera \textit{et al.} \cite{mosq} suggested a process that 
would lead to the formation of charged black holes. In this context, it 
will be of interest the study of possible signatures of 
charged black holes. Zakharov \textit{et al.} \cite{zakharov} proposed that 
the charge of the Galactic center black hole could be measured by a future 
space based radio interferometer (RADIOASTRON), using gravitational 
lensing. As mentioned above, Reissner--Nordstr\"{o}m black holes 
acting as gravitational lenses were studied in Ref. \cite{eiroto}, and 
Born--Infeld electrodynamics was suggested as a possible alternative to 
Maxwell electrodynamics by recent developments of low energy string theory, 
so the analysis of gravitational lensing properties of Einstein--Born--Infeld 
black holes will be the natural next step. The purpose 
of this article is a comprehensive study of Einstein Born--Infeld 
black holes as gravitational lenses and the comparison of the results 
obtained with those corresponding to Reissner--Nordstr\"{o}m geometry. The 
paper is organized as follows. In Sec. 2, the Einstein--Born--Infeld black 
holes are reviewed. In  Sec. 3, the expression for the deflection angle 
is found by means of the strong deflection limit. In Sec. 4, the lens 
equation is used to obtain the positions and magnifications of the 
relativistic images. In Sec. 5, as numerical examples, the lensing observables 
for the supermassive Galactic center black hole and for a small size black 
hole placed at the Galactic halo are calculated. Finally, in Sec. 6, a 
discussion of the results is made.

\section{Einstein--Born--Infeld black holes}

The action of Einstein gravity coupled to Born--Infeld electrodynamics\footnote
{Throughout the paper, units such as $G=c=4\pi\epsilon_0=
(4\pi )^{-1} \mu _{0}=1$ are adopted, and the signature of the metric is 
taken ($-+++$).} has the form
\begin{equation}
S=\int dx^{4}\sqrt{-g}\left ( \frac{R}{16\pi } +L_{BI}\right) ,
\label{bi0a}
\end{equation}
with
\begin{equation}
L_{BI}=\frac{1}{4\pi b^{2}}\left( 1-\sqrt{1+
\frac{1}{2}F_{\sigma \nu }F^{\sigma \nu }b^{2}-
\frac{1}{4}\mbox{} ^{*}F_{\sigma \nu }F^{\sigma \nu }b^{4}}\right) ,
\label{bi0b}
\end{equation}
where $g$ is the determinant of the metric tensor, $R$ is 
the scalar of curvature, $F_{\sigma \nu }=\partial _{\sigma }A_{\nu }
-\partial _{\nu }A_{\sigma }$ is the electromagnetic tensor,  
$\mbox{} ^{*}F_{\sigma \nu } =\tfrac{1}{2}\sqrt{-g}\, \varepsilon _{\alpha 
\beta \sigma \nu }F^{\alpha \beta }$ is the Hodge dual of $F_{\sigma \nu }$ 
(with $\varepsilon _{\alpha \beta \sigma \nu }$ the Levi--Civita symbol) and 
$b$ is a parameter that indicates how much Born--Infeld and Maxwell 
electrodynamics differ. In the context of string theory $b$ is related with 
the string tension. For $b\rightarrow 0$ the Einstein--Maxwell action is 
recovered. The field equations can be obtained by varying the action 
(\ref{bi0a}) with respect to the metric $g_{\sigma \nu }$ and the 
electromagnetic potential $A_{\nu }$. These field equations have spherically 
symmetric black hole solutions \cite {gibbons} given by 
\begin{equation}
ds^{2}=-\psi (r)dt^{2}+\psi (r)^{-1}dr^{2}+r^{2}d\Omega ^{2},
\label{bi1}
\end{equation}
with 
\begin{equation}
\psi (r)=1-\frac{2M}{r}+\frac{2}{b^{2}r}\int^{\infty}_{r}
\left( \sqrt {x^{4}+b^{2}Q^{2}}-x^{2}\right) dx,
\label{bi2}
\end{equation} 
\begin{equation}
D(r)=\frac{Q_{E}}{r^{2}},
\label{bi2a}
\end{equation}
\begin{equation}
B(r)=Q_{M}\sin \theta,
\label{bi2b}
\end{equation}
where $M$ is the ADM mass, $Q^{2}=Q_{E}^{2}+Q_{M}^{2}$ is the sum of the 
squares of the electric $Q_{E}$ and magnetic $Q_{M}$ charges, $B(r)$ and 
$D(r)$ are the  magnetic and the electric inductions in the local 
orthonormal frame. In the limit $b\rightarrow 0$, the Reissner--Nordstr\"{o}m 
metric is obtained. The metric (\ref{bi1}) is also asymptotically 
Reissner--Norsdtr\"{o}m for large values of $r$. With the units adopted above, 
$M$, $Q$ and $b$ have dimensions of length. The metric function $\psi (r)$ 
can be expressed in the form
\begin{equation}
\psi (r)=1-\frac{2M}{r}+\frac{2}{3b^{2}}\left\{ r^{2}-\sqrt{r^{4}+b^{2}Q^{2}}+
\frac{\sqrt{|bQ|^{3}}}{r}F\left[ \arccos\left( \frac{r^{2}-|bQ|}{r^{2}+|bQ|}
\right) ,\frac{\sqrt{2}}{2}\right]\right\} ,
\label{bi3}
\end{equation}
where $F(\gamma ,k)$ is the elliptic integral of the first kind\footnote{
$F(\gamma , k)=\int _{0}^{\gamma }(1-k^{2}\sin ^{2}\phi )^{-1/2}d\phi =
\int _{0}^{\sin \gamma }[(1-z^{2})(1-k^{2}z^{2})]^{-1/2}dz $ 
\cite{gradshteyn}}. As in Schwarzschild and Reissner--Norsdtr\"{o}m cases, the 
metric (\ref{bi1}) has a singularity at $r=0$ \cite{breton}. The 
zeros of $\psi (r)$ determine the position of the horizons, which have to 
be obtained numerically. For a given value of $b$, when the charge is small, 
$0\le |Q|/M\le \nu _{1}$, the function $\psi (r)$ has one zero and there 
is a regular event horizon. For intermediate values of charge, 
$\nu _{1}<|Q|/M<\nu _{2}$, $\psi (r)$ has two zeros, so there are, as in the 
Reissner--Nordstr\"{o}m geometry, an inner horizon and an outer regular event 
horizon. When $|Q|/M=\nu _{2}$, there is one degenerate horizon. Finally, if 
the values of charge are large, $|Q|/M>\nu _{2}$, the function $\psi (r)$ has 
no zeros and a naked singularity is obtained. The values of $|Q|/M$ where the 
number of horizons change, $\nu _{1}=(9|b|/M)^{1/3}[F(\pi ,\sqrt{2}/2]^{-2/3}$ 
and $\nu _{2}$, which should be calculated numerically from the condition 
$\psi (r_{h})=\psi '(r_{h})=0$, are increasing functions of $|b|/M$. In 
the Reissner--Nordstr\"{o}m limit ($b\rightarrow 0$) it is easy to see that 
$\nu _{1}=0$ and $\nu _{2}=1$.\\

The paths of photons in nonlinear electrodynamics are not null geodesics of 
the background geometry. Instead, they follow null geodesics of an effective 
metric \cite{pleb} generated by the self-interaction of the electromagnetic 
field, which depends on the particular nonlinear theory considered. In 
Einstein gravity coupled to Born--Infeld electrodynamics the effective 
geometry for photons is given by \cite{breton}:
\begin{equation}
ds^{2}_{eff}=-\omega (r)^{1/2}\psi (r)dt^{2}+\omega (r)^{1/2}\psi (r)^{-1}
dr^{2}+\omega (r)^{-1/2}r^{2}d\Omega ^{2}, 
\label{bi4}
\end{equation}
where 
\begin{equation}
\omega (r)=1+\frac{Q^{2}b^{2}}{r^{4}}.
\label{bi5}
\end{equation}
Then, to calculate the deflection angle for photons passing near the black 
holes considered in the present work, it is necessary to use the effective 
metric (\ref{bi4}) instead of the background metric (\ref{bi1}). The horizon 
structure of the effective metric is the same that of metric (\ref{bi1}), but 
the trajectories of photons are different.

\section{Deflection angle}

For a spherically symmetric black hole with a metric of the form 
\begin{equation}
ds^{2}=-f(r)dt^{2}+g(r)dr^{2}+h(r)d\Omega ^{2},
\label{sl1}
\end{equation}
the radius of the event horizon $r_{h}$ is given by the greatest positive 
root of the equation $f(r)=0$, and the radius of the photon sphere $r_{ps}$ 
by the greatest positive solution of the equation $f(r)h'(r)=f'(r)h(r)$,
where the prime means the derivative respect to the radial coordinate $r$. 
The deflection angle for a photon coming from infinite is given by 
\cite{weinberg}
\begin{equation}
\alpha (r_{0})=I(r_{0})-\pi ,
\label{sl4}
\end{equation}
where $r_0$ is the closest approach distance and
\begin{equation}
I(r_{0})=\int_{r_{0}}^{\infty }2\left[ \frac {g(r)}{h(r)}\right] ^{1/2}
\left[ \frac {h(r)f(r_{0})}{h(r_{0})f(r)}-1\right] ^{-1/2}dr.
\label{sl5}
\end{equation}
For the effective metric (\ref{bi4}) the integral is
\begin{equation}
I(r_{0})=2\int_{r_{0}}^{\infty }\frac{r_{0}\omega (r)}
{r\sqrt{\omega (r_{0})\psi (r_{0})r^{2}-r_{0}^{2}\omega (r)\psi (r)}}dr,
\label{sl6}
\end{equation}
thus, defining $\chi (r)=\omega (r)\psi (r)$ and making the substitution 
$z=r_{0}/r$, it takes the form
\begin{equation}
I(r_{0})=2\int_{0}^{1}\frac{\omega (r_{0}/z)}
{\sqrt{\chi (r_{0})-\chi (r_{0}/z)z^{2}}}dz.
\label{sl7}
\end{equation}
The functions inside the integral $\omega (r_{0}/z)$ and 
$\chi (r_{0})-\chi (r_{0}/z)z^{2}$ can be expanded in powers of $1-z$: 
\begin{equation}
\omega (r_{0}/z)=1+\frac{Q^{2}b^{2}}{r_{0}^{4}}-4\frac{Q^{2}b^{2}}{r_{0}^{4}}
(1-z)+O(1-z)^{2},
\label{sl8a}
\end{equation}
\begin{equation}
\chi (r_{0})-\chi (r_{0}/z)z^{2}=\gamma_{1}(r_{0})(1-z)+
\gamma_{2}(r_{0})(1-z)^{2}+O(1-z)^{3},
\label{sl8b}
\end{equation}
where $\gamma_{1}(r_{0})=2\chi (r_{0})-r_{0}\chi '(r_{0})$ and 
$\gamma_{2}(r_{0})=-\chi (r_{0})+r_{0}\chi '(r_{0})-(1/2)r_{0}^{2}
\chi ''(r_{0})$. For the effective metric (\ref{bi4}) the photon sphere 
radius $r_{ps}$ is given by the greatest positive solution of the equation 
$2\chi (r)-r\chi '(r)=0$, which should be solved numerically. When 
$r_{0}>r_{ps}$, $\gamma_{1}(r_{0})\neq 0$ and the leading term inside the 
integral in Eq. (\ref{sl7}) is proportional to $1/\sqrt{(1-z)}$, so it 
converges. As $\gamma_{1}(r_{ps})=0$, for $r_{0}=r_{ps}$ the leading term goes 
as $1/(1-z)$ and the integral has a logarithmic divergence. Then, following 
Ref. \cite{bozza2}, it is convenient to separate $I(r_{0})$ as a sum of two 
parts:
\begin{equation}
I(r_{0})=I_{D}(r_{0})+I_{R}(r_{0}),
\label{sl9}
\end{equation}
with
\begin{equation}
I_{D}(r_{0})=2\int_{0}^{1}\frac{r_{0}^{4}+Q^{2}b^{2}}
{r_{0}^{4}\sqrt{\gamma_{1}(r_{0})(1-z)+\gamma_{2}(r_{0})(1-z)^{2}}}dz,
\label{sl10}
\end{equation}
and
\begin{equation}
I_{R}(r_{0})=2\int_{0}^{1}\left[ \frac{\omega (r_{0}/z)}
{\sqrt{\chi (r_{0})-\chi (r_{0}/z)z^{2}}}-\frac{r_{0}^{4}+Q^{2}b^{2}}
{r_{0}^{4}\sqrt{\gamma_{1}(r_{0})(1-z)+\gamma_{2}(r_{0})(1-z)^{2}}}\right] dz.
\label{sl11}
\end{equation} 
The integral $I_{D}(r_{0})$, which diverges for $r_{0}=r_{ps}$, can be 
calculated exactly to give
\begin{equation}
I_{D}(r_{0})=-\frac{2\left( r_{0}^{4}+Q^{2}b^{2}\right) }
{r_{0}^{4}\sqrt{\gamma_{2}(r_{0})}} \ln \frac{\gamma_{1}(r_{0})}
{\left[ \sqrt{\gamma_{1}(r_{0})+\gamma_{2}(r_{0})}+\sqrt{\gamma_{2}(r_{0})}
\right]^{2}},
\label{sl12}
\end{equation} 
and for $r_0$ close to $r_{ps}$ it takes the form
\begin{equation}
I_{D}(r_{0})=-\frac{\sqrt{8}\left( r_{ps}^{4}+Q^{2}b^{2}\right) }
{r_{ps}^{4}\sqrt{2\chi (r_{ps})-r_{ps}^{2}\chi ''(r_{ps})}}
\left[ \ln\left( \frac{r_{0}}{r_{ps}}-1 \right) -\ln 2 \right] 
+O(r_{0}-r_{ps}),
\label{sl13}
\end{equation}
where it was used that $2\chi (r_{ps})-r\chi '(r_{ps})=0$. The integral 
$I_{R}(r_{0})$ is the original integral $I(r_{0})$ with the 
divergence subtracted, so it converges when $r_{0}=r_{ps}$, and it can be 
replaced by $I_{R}(r_{0})=I_{R}(r_{ps})+O(r_{0}-r_{ps})$.\\
 
The deflection angle, which diverges for $r_0=r_{ps}$, is thus given 
in the strong deflection limit by
\begin{equation}
\alpha(r_{0})=-a_1\ln\left(\frac{r_{0}}{r_{ps}}-1\right)+a_2+O(r_{0}-r_{ps}),
\label{sl14}
\end{equation}
where
\begin{equation}
a_{1}=\frac{\sqrt{8}\left( r_{ps}^{4}+Q^{2}b^{2}\right) }
{r_{ps}^{4}\sqrt{2\chi (r_{ps})-r_{ps}^{2}\chi ''(r_{ps})}},
\label{sl15}
\end{equation}
and
\begin{equation}
a_{2}=-\pi +a_{D}+a_{R},
\label{sl16}
\end{equation}
with
\begin{equation}
a_{D}=a_{1}\ln{2},
\label{sl17}
\end{equation}
and
\begin{equation}
a_{R}=I_{R}(r_{ps})=2\int_{0}^{1}\left[\frac{\omega (r_{ps}/z)}
{\sqrt{\chi (r_{ps})-\chi (r_{ps}/z)z^{2}}}-
\frac{\sqrt{2}\left( r_{ps}^{4}+Q^{2}b^{2}\right) }
{(1-z)r_{ps}^{4}\sqrt{2\chi (r_{ps})-r_{ps}^{2}\chi ''(r_{ps})}}\right]dz.
\label{sl18}
\end{equation}
\begin{figure}[t!]
\begin{center}
\vspace{0cm}
\includegraphics[width=17cm]{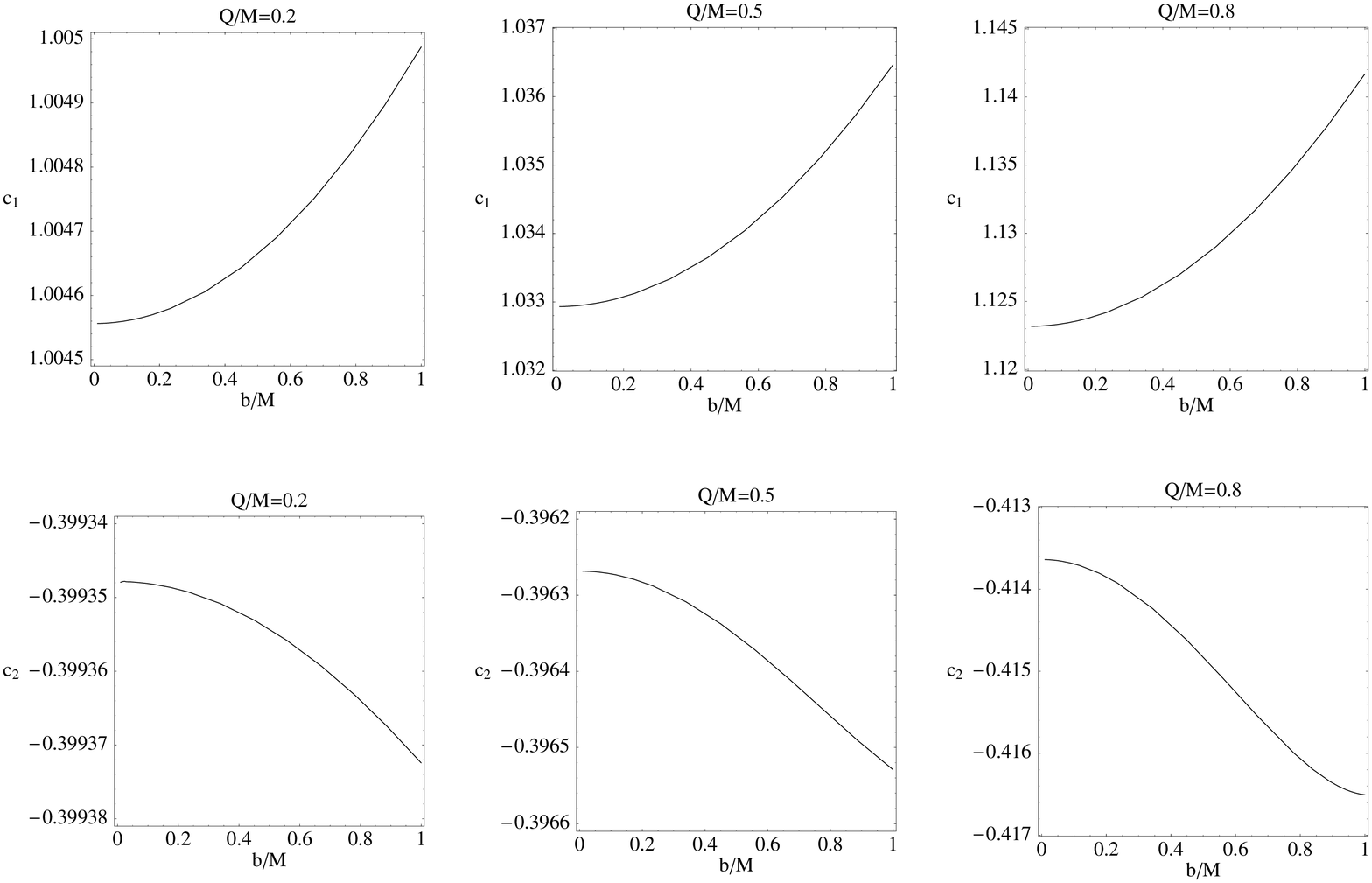}
\vspace{-1cm}
\end{center} 
\vspace{0cm} 
\caption{The strong deflection limit coefficients $c_{1}$ (upper panel) and 
$c_{2}$ (lower panel) as functions of the Born--Infeld parameter $b$ for 
different values of the charge $Q$. When $b=0$ the coefficients corresponding 
to Reissner--Nordstr\"{o}m black holes are obtained. The Schwarzschild values 
are $c_{1}^{Schw}=1$ and $c_{2}^{Schw}=\ln [216(7-4\sqrt{3})]-\pi \approx
-0.40023$.}
\label{fig1}
\end{figure}
As it happens in the Reissner-Nordstr\"{o}m case, when $Q\neq0$ it is not 
possible to calculate this integral in an exact form. It can be obtained 
approximately, by means of a numerical treatment or making a power expansion 
in $Q$. The impact parameter $u$, defined as the perpendicular distance from 
the black hole to the asymptotic path at infinite, is more easily related with 
the lensing angles than the closest approach distance $r_{0}$. Following 
Ref. \cite{weinberg}, $u=[h(r_{0})/f(r_{0})]^{1/2}$, which in our case gives 
$u=r_{0}/\sqrt{\chi (r_{0})}$. Making a second order Taylor expansion around 
$r_{0}=r_{ps}$, it takes the form
\begin{equation}
u=u_{ps}+\frac{2\chi (r_{ps})-r_{ps}^{2}\chi ''(r_{ps})}{4r_{ps}
[\chi (r_{ps})]^{3/2}}(r_{0}-r_{ps})^{2} +O(r_{0}-r_{ps})^{3},
\label{sl20}
\end{equation}
where $u_{ps}=r_{ps}/\sqrt{\chi (r_{ps})}$ is the critical impact parameter. 
Inverting Eq. (\ref{sl20}):
\begin{equation}
\frac{r_{0}}{r_{ps}}-1=\left[ \frac{2\chi (r_{ps})-r_{ps}^{2}\chi ''(r_{ps})}
{4\chi (r_{ps})}\right] ^{-1/2}
\left( \frac{u}{u_{ps}}-1\right)^{1/2}.
\label{sl21}
\end{equation}
Then the deflection angle can be obtained as a function of the impact 
parameter $u$:
\begin{equation}
\alpha (u)=-c_{1}\ln \left( \frac{u}{u_{ps}}-1 \right) +c_{2}+O(u-u_{ps}), 
\label{sl22}
\end{equation}
with $c_{1}=a_{1}/2$ and
\begin{equation} 
c_{2}=\frac{a_{1}}{2}\ln{\frac{2\chi (r_{ps})-r_{ps}^{2}\chi ''(r_{ps})}
{4\chi (r_{ps})}}+a_{2}. 
\label{sl23}
\end{equation}
Eq. (\ref{sl22}) represents the strong deflection limit approximation of the 
deflection angle as a function of the impact parameter. Photons with an 
impact parameter slightly greater than the critical value $u_{ps}$ will spiral 
out, eventually reaching the observer after one or more turns around the black 
hole. In this case, the strong deflection limit gives a good approximation for 
the deflection angle. Those photons whose impact parameter is smaller than 
$u_{ps}$ will spiral into the black hole, not reaching any observer outside 
the photon sphere. The coefficients $c_{1}$ and $c_{2}$ (obtained numerically) 
are plotted as functions of the Born--Infeld parameter $b$ in 
Fig. \ref{fig1}. For a given non null value of charge, $c_{1}$ increases and 
$c_{2}$ decreases (becomes more negative) with $b$, thus the 
Einstein--Born--Infeld black holes have larger $c_{1}$ and smaller $c_{2}$ 
than their Reissner--Norsdtr\"{o}m counterparts with the same charge. The 
differences between these geometries grow as $|Q|$ increases. When the 
charge is zero, the Schwarzschild geometry is recovered, with the strong field 
limit coefficients given by $c_{1}^{Schw}=1$ and 
$c_{2}^{Schw}=\ln [216(7-4\sqrt{3})]-\pi \approx 0.40023$.

\section{Positions and magnifications of the relativistic images}

The lens geometry consists of a point source of light (s), a black hole, which 
it is called the lens (l), and the observer (o), with three possible 
configurations of them. The configuration where the lens between the observer 
and the source is named standard lensing (SL). Those corresponding to  
the source between the observer and the lens (RLI), or the observer between 
the source and the lens (RLII) are called retrolensing. The line 
joining the observer and the lens define the optical axis and the background 
space-time is asymptotically flat, with both the observer and the source in 
the flat region. The angular positions, seen from the observer, of the source 
and the images are, respectively, $\beta $ (taken positive without losing 
generality) and $\theta $. The observer-source ($d_{os}$), observer-lens 
($d_{ol}$) and the lens-source ($d_{ls}$) distances (measured along the 
optical axis) are taken much greater than the horizon radius. The lens 
equation is given by
\begin{equation}
\tan \beta =\tan \theta -c_{3}\left[ \tan (\alpha -\theta)
+\tan \theta \right] ,
\label{pm1}
\end{equation}
where $c_{3}=d_{ls}/d_{os}$ (SL) \cite{virbha1}, $c_{3}=d_{os}/d_{ol}$ (RLI) 
\cite{eitor} or $c_{3}=d_{os}/d_{ls}$ (RLII) \cite{eiroa}, depending on the 
configuration considered. The lensing effects are more important when the 
objects are highly aligned, so the analysis will be restricted to this 
case\footnote{for a unified treatment of standard lensing, retrolensing and 
intermediate situations in the strong deflection limit, see Ref. 
\cite{bozman2}}, in which the angles $\beta $ and $\theta $ are small, and 
$\alpha $ is close to an even multiple of $\pi $ for standard lensing or to an 
odd multiple of $\pi $ for retrolensing. In the standard lensing case, when 
$\beta \neq 0$ two weak field primary and secondary images, which will 
be not considered here, and two infinite sets of point relativistic images are 
formed. The first set of relativistic images have a deflection angle that can 
be written as $\alpha =2n\pi +\Delta \alpha _{n}$, with $n\in \mathbb{N}$ and 
$0<\Delta \alpha _{n}\ll 1$. Then, the lens equation can be simplified:
\begin{equation}
\beta =\theta -c_{3}\Delta \alpha _{n}.
\label{pm2}
\end{equation}
To obtain the other set of images, it should be taken 
$\alpha =-2n\pi -\Delta \alpha _{n}$, so $\Delta \alpha _{n}$ must
be replaced by $-\Delta \alpha _{n}$ in Eq. (\ref{pm2}). From the lens 
geometry it is easy to see that $u=d_{ol}\sin \theta $, which can be 
approximated to first order in $\theta $ by $u=d_{ol}\theta $, so the 
deflection angle given by Eq. (\ref{sl22}) can be written as a function 
of $\theta $:
\begin{equation}
\alpha (\theta )=-c_{1}\ln \left( \frac{d_{ol}\theta }{u_{ps}}-1 \right) 
+c_{2}. 
\label{pm4}
\end{equation}
Then, inverting Eq. (\ref{pm4}) to obtain $\theta (\alpha )$
\begin{equation}
\theta (\alpha )=\frac{u_{ps}}{d_{ol}}\left[ 1+e^{(c_{2}-\alpha )/c_{1}}
\right] ,
\label{pm5}
\end{equation}
and making a first order Taylor expansion around $\alpha =2n\pi $, the angular 
position of the $n$-th image is
\begin{equation}
\theta _{n}=\theta ^{0}_{n}-\zeta _{n}\Delta \alpha _{n},
\label{pm6}
\end{equation}
with
\begin{equation}
\theta ^{0}_{n}=\frac{u_{ps}}{d_{ol}}\left[ 1+e^{(c_{2}-2n\pi )/c_{1}}
 \right] ,
\label{pm7}
\end{equation}
and
\begin{equation}
\zeta _{n}=\frac{u_{ps}}{c_{1}d_{ol}}e^{(c_{2}-2n\pi )/c_{1}}.
\label{pm8}
\end{equation}
From Eq. (\ref{pm2}), $\Delta \alpha _{n}=(\theta _{n}-\beta )/c_{3}$, so 
replacing it in Eq. (\ref{pm6}) gives
\begin{equation}
\theta _{n}=\theta ^{0}_{n}-\frac{\zeta _{n}}{c_{3}}(\theta _{n}-\beta ),
\label{pm10}
\end{equation}
which can be expressed in the form
\begin{equation}
\theta _{n}=\left( 1+\frac{\zeta _{n}}{c_{3}}\right) ^{-1}\left( 
\theta ^{0}_{n}+\frac{\zeta _{n}}{c_{3}}\beta \right) ,
\label{pm12}
\end{equation}
then, using that $0<\zeta _{n}/c_{3}\ll 1$ and keeping only the first order 
term in $\zeta _{n}/c_{3}$, the angular positions of the images can be 
approximated by
\begin{equation}
\theta _{n}=\theta ^{0}_{n}+\frac {\zeta _{n}}{c_{3}}(\beta -\theta ^{0}_{n}).
\label{pm14}
\end{equation}
The second term in Eq. (\ref{pm14}) is a small correction on 
$\theta ^{0}_{n}$, so all images lie very close to $\theta ^{0}_{n}$. With a 
similar treatment, the other set of relativistic images have angular 
positions 
\begin{equation}
\theta _{n}=-\theta ^{0}_{n}+\frac {\zeta _{n}}{c_{3}}(\beta +\theta ^{0}_{n}).
\label{pm15}
\end{equation}
In the case of perfect alignment ($\beta =0$), instead of point images an 
infinite sequence of concentric rings is obtained, with angular radius
\begin{equation}
\theta ^{E}_{n}=\left( 1-\frac {\zeta _{n}}{c_{3}}\right) \theta ^{0}_{n},
\label{pm16}
\end{equation}
which are usually called Einstein rings.\\

It is a well known result that gravitational lensing conserves surface 
brightness \cite{schne}, so the ratio of the solid angles subtended by the 
image and the source gives the magnification of the $n$-th image:
\begin{equation}
\mu _{n}=\left| \frac{\sin \beta }{\sin \theta _{n}}
\frac{d\beta }{d\theta _{n}}\right|^{-1}\approx 
\left| \frac{\beta }{\theta _{n}} \frac{d\beta }{d\theta _{n}}\right|^{-1},
\label{pm17}
\end{equation}
which, using Eq. (\ref{pm14}), leads to
\begin{equation}
\mu _{n}=\frac{1}{\beta}\left[ \theta ^{0}_{n}+
\frac {\zeta _{n}}{c_{3}}(\beta - \theta ^{0}_{n})\right]
\frac {\zeta _{n}}{c_{3}},
\label{pm18}
\end{equation}
that can be approximated to first order in $\zeta _{n}/c_{3}$ by
\begin{equation}
\mu _{n}=\frac{1}{\beta}\frac{\theta ^{0}_{n}\zeta _{n}}{c_{3}}.
\label{pm19}
\end{equation}
The same expression is obtained for the other set of relativistic images. 
The first image is the brightest one, and the magnifications
decrease exponentially with $n$. For retrolensing, the same equations 
for the positions and magnifications of the relativistic images apply, 
with $2n$ replaced by $2n-1$ in the expressions of $\theta ^{0}_{n}$ and 
$\zeta _{n}$. The magnifications of the images are greater in retrolensing 
configurations. In all cases, the magnifications are proportional to 
$(u_{ps}/d_{ol})^{2}$, which is a very small factor. Then, the relativistic 
images are very faint, unless $\beta $ has values close to zero, 
i.e. nearly perfect alignment. For $\beta =0$, the magnification becomes 
infinite, and the point source approximation breaks down, so an extended 
source analysis is needed. The magnification of the images for an extended 
source is obtained by integrating over its luminosity profile:
\begin{equation}
\mu _{n}^{ext}=\frac{\int\!\!\int_{S}\mathcal{I}\mu _{p}dS}{\int\!\!\int_{S}
\mathcal{I}dS}, \label{pm22}
\end{equation}
where $\mathcal{I}$ is the surface intensity distribution of the source and
$\mu _{p}$ is the magnification corresponding to each point of the source.
When the source is an uniform disk $D(\beta _{c},\beta _{s})$, with angular 
radius $\beta _{s}$ and centered in $\beta _{c}$ (taken positive), 
Eq. (\ref{pm22}) takes the form
\begin{equation}
\mu _{n}^{ext}=\frac{\int\!\!\int_{D(\beta _{c},\beta _{s})}\mu _{p} dS}
{\pi \beta _{s}^{2}}.
\label{pm23}
\end{equation}
So, using Eq. (\ref{pm19}), the magnification of the relativistic $n$-th 
image for an extended uniform source is
\begin{equation}
\mu _{n}^{ext}=\frac{I}{\pi \beta _{s}^{2}}\frac {\theta ^{0}_{n}\zeta _{n}}
{c_{3}},
\label{pm24}
\end{equation}
with 
\begin{equation}
I=2\left[ (\beta _{s}+\beta _{c})
E\left(\frac{2\sqrt{\beta _{s}\beta _{c}}}{\beta _{s}+\beta _{c}}\right)
+(\beta _{s}-\beta _{c})K\left(\frac{2\sqrt{\beta _{s}\beta _{c}}}
{\beta _{s}+\beta _{c}}\right) \right] ,
\label{pm25}
\end{equation}
where $K(k)$ and $E(k)$ are respectively, complete elliptic integrals of the 
first\footnote{$K(k)=\int _{0}^{\pi/2}(1-k^{2}\sin ^{2}\phi )^{-1/2}d\phi =
\int _{0}^{1}[(1-z^{2})(1-k^{2}z^{2})]^{-1/2}dz $ \cite{gradshteyn}} and 
second\footnote{$E(k)=\int_{0}^{\pi /2}\left( 1-k^2\sin ^{2}\phi \right) ^{1/2}
d\phi =\int _{0}^{1}(1-z^{2})^{-1/2}(1-k^{2}z^{2})^{1/2}dz $ 
\cite{gradshteyn}} kind. Finite magnifications are always obtained from Eq. 
(\ref{pm24}), even in the case of complete alignment.\\ 

The total magnification, taking into account both sets of images, is 
$\mu =2\sum\limits_{n=1}^{\infty }\mu _{n}$ which for a point source, using 
Eqs. (\ref{pm7}), (\ref{pm8}) and (\ref{pm19}), gives
\begin{equation}
\mu =\frac {1}{\beta}\frac {2u_{ps}^{2}}{d_{ol}^{2}c_{1}c_{3}}
\frac{e^{c_{2}/c_{1}}\left( 1+e^{c_{2}/c_{1}}+e^{2\pi /c_{1}}\right) }
{e^{4\pi /c_{1}}-1}
\label{pm20}
\end{equation}
for standard lensing, and 
\begin{equation}
\mu =\frac {1}{\beta}\frac {2u_{ps}^{2}}{d_{ol}^{2}c_{1}c_{3}}
\frac{e^{(c_{2}+\pi )/c_{1}}\left[ 1+e^{(c_{2}+\pi )/c_{1}}+e^{2\pi /c_{1}}
\right] }{e^{4\pi /c_{1}}-1}
\label{pm21}
\end{equation}
for retrolensing. When the source is extended, the total magnification is
given by
\begin{equation}
\mu ^{ext}=\frac{I}{\pi \beta _{s}^{2}}\frac {2u_{ps}^{2}}
{d_{ol}^{2}c_{1}c_{3}}\frac{e^{c_{2}/c_{1}}
\left( 1+e^{c_{2}/c_{1}}+e^{2\pi /c_{1}}\right) }{e^{4\pi /c_{1}}-1}
\label{pm26}
\end{equation}
for standard lensing, and
\begin{equation}
\mu ^{ext}=\frac{I}{\pi \beta _{s}^{2}}\frac {2u_{ps}^{2}}
{d_{ol}^{2}c_{1}c_{3}}\frac{e^{(c_{2}+\pi )/c_{1}}
\left[ 1+e^{(c_{2}+\pi )/c_{1}}+e^{2\pi /c_{1}}\right] }{e^{4\pi /c_{1}}-1}
\label{pm27}
\end{equation}
for retrolensing.

\section{Examples}

In this Section, only intended to give some feeling of the numbers involved, 
the model is applied to the supermassive Galactic center black hole and to a 
low mass black hole at the Galactic halo. The first image, with angular 
position $\theta _{1}$, is the outermost one, and the others approach to the 
limiting value $\theta _{\infty}=u_{ps}/d_{ol}$ as $n$ increases. The lensing 
observables defined by Bozza \cite{bozza2}:
\begin{equation}
s=\theta _{1}-\theta _{\infty }
\label{g1a}
\end{equation}
and
\begin{equation}
r=\frac{\mu _{1}}{\sum\limits_{n=2}^{\infty }\mu _{n}},
\label{g1b}
\end{equation}
are useful when the outermost image can be resolved from the rest. The 
observable $s$ represents the angular separation between the first image and 
the limiting value of the succession of images, and $r$ is the ratio between 
the flux of the first image and the sum of the fluxes of the other images. For 
high alignment, they can be approximated by
\begin{equation}
s_{\scriptscriptstyle{SL}}=\theta_{\infty }e^{(c_{2}-2\pi )/c_{1}},
\label{g2a}
\end{equation}
\begin{equation}
r_{\scriptscriptstyle{SL}}=e^{2\pi /c_{1}}+e^{c_{2}/c_{1}}-1,
\label{g2b}
\end{equation} 
for standard lensing, and by
\begin{equation}
s_{\scriptscriptstyle{RL}}=\theta_{\infty }e^{(c_{2}-\pi )/c_{1}},
\label{g3a}
\end{equation}
\begin{equation}
r_{\scriptscriptstyle{RL}}=e^{2\pi /c_{1}}+e^{(c_{2}+\pi )/c_{1}}-1,
\label{g3b}
\end{equation} 
for retrolensing. The strong deflection limit coefficients $c_{1}$ and $c_{2}$ 
can be obtained by measuring $\theta_{\infty }$, $s$ and $r$ and inverting 
Eqs. (\ref{g2a}) and (\ref{g2b}) (SL) or Eqs. (\ref{g3a}) and (\ref{g3b}) 
(RL). Then, their values can be compared with those predicted by the 
theoretical models to identify the nature of the black hole lens.\\

\begin{figure}[t!]
\begin{center}
\vspace{0cm}
\includegraphics[width=17cm]{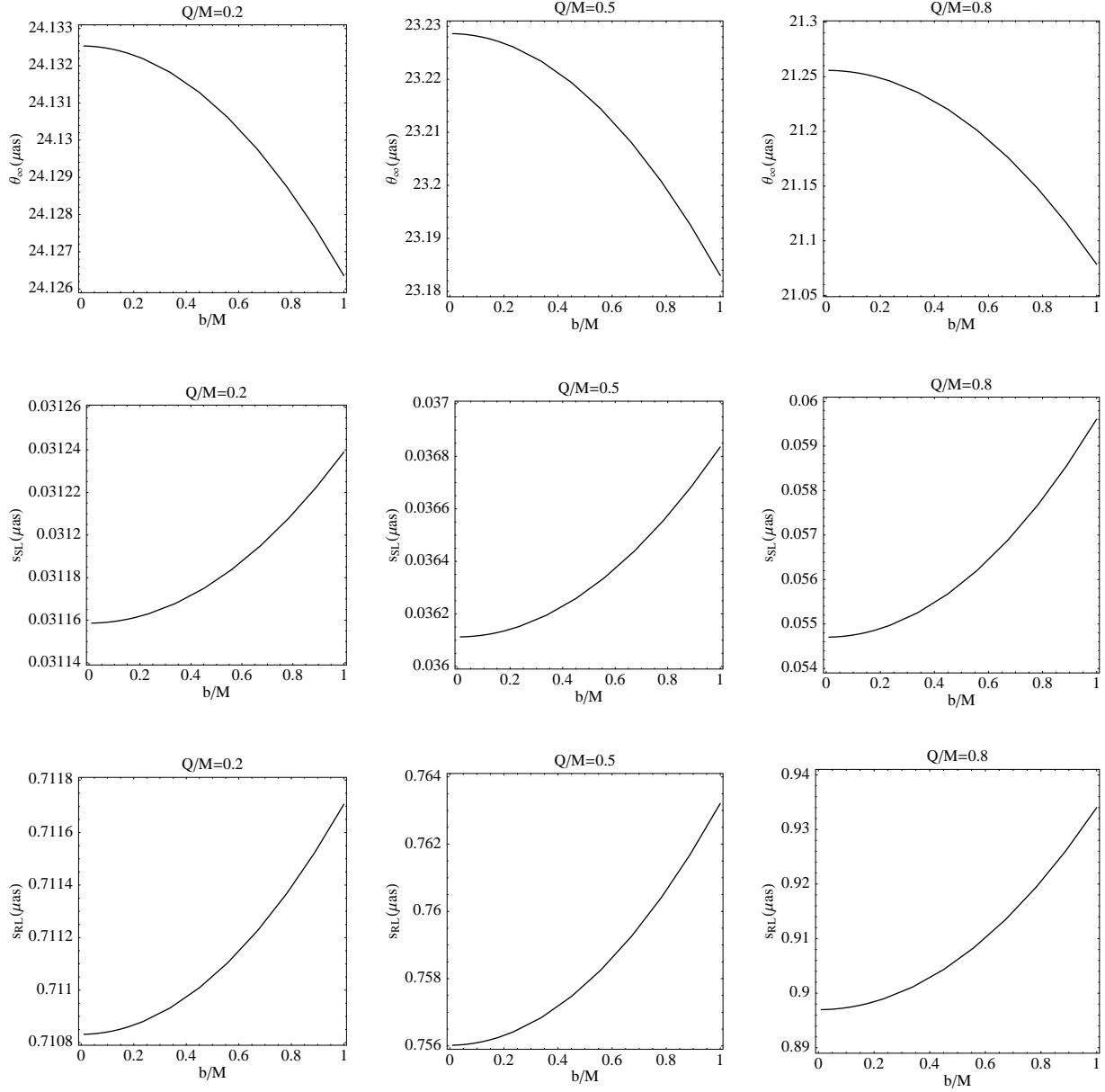}
\vspace{-1cm}
\end{center} 
\vspace{0cm} 
\caption{Gravitational lensing by the Galactic center black hole. The values 
in microarcseconds ($\mu \mathrm{as}$) of $\theta _{\infty }$ (upper panel), 
$s_{\scriptscriptstyle{SL}}$ (center) and $s_{\scriptscriptstyle{RL}}$ (lower 
panel) are plotted as functions of the Born--Infeld parameter $b$ for 
different values of the charge $Q$. When $b=0$ the results corresponding to 
Reissner--Nordstr\"{o}m geometry are obtained. The Schwarzschild black hole 
values are $\theta _{\infty }^{Schw}=24.296\, \mu \mathrm{as}$, 
$s_{\scriptscriptstyle{SL}}^{Schw}=0.03041\, \mu \mathrm{as}$ and 
$s_{\scriptscriptstyle{RL}}^{Schw}=0.7036\, \mu \mathrm{as}$.}
\label{fig2}
\end{figure}
\begin{figure}[t!]
\begin{center}
\vspace{0cm}
\includegraphics[width=17cm]{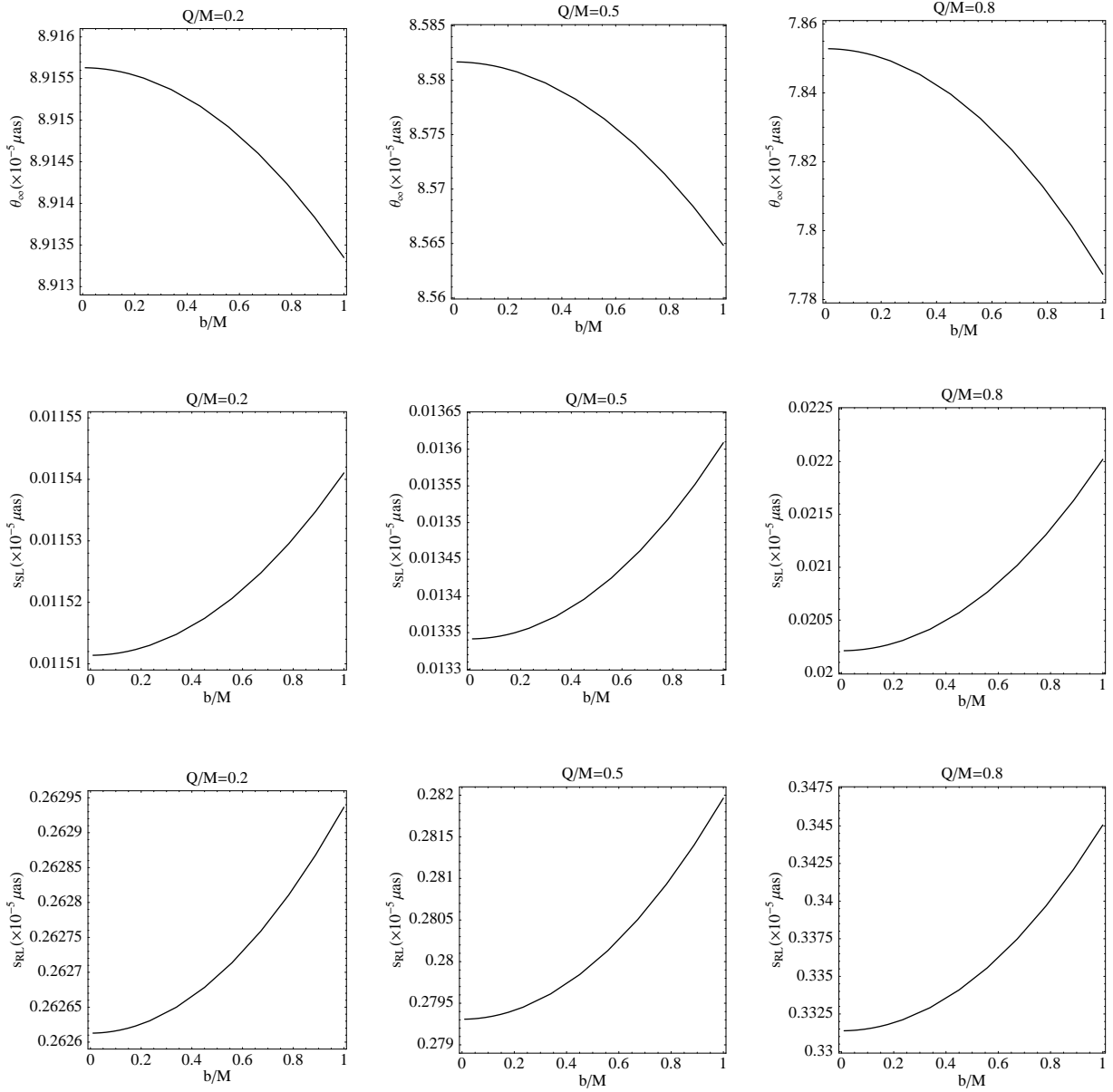}
\vspace{-1cm}
\end{center} 
\vspace{0cm} 
\caption{Gravitational lensing by a low mass black hole in the Galactic 
halo. The values in microarcseconds ($\mu \mathrm{as}$) of 
$\theta _{\infty }$ (upper panel), $s_{\scriptscriptstyle{SL}}$ (center) 
and $s_{\scriptscriptstyle{RL}}$ (lower panel) are plotted as 
functions of the Born--Infeld parameter $b$ for different values 
of the charge $Q$. When $b=0$ the results corresponding to 
Reissner--Nordstr\"{o}m geometry are obtained. The Schwarzschild black hole 
values are $\theta _{\infty }^{Schw}=24.296\, \mu \mathrm{as}$, 
$s_{\scriptscriptstyle{SL}}^{Schw}=0.03041\, \mu \mathrm{as}$ and 
$s_{\scriptscriptstyle{RL}}^{Schw}=0.7036\, \mu \mathrm{as}$.}
\label{fig3}
\end{figure}
\begin{figure}[t!]
\begin{center}
\vspace{0cm}
\includegraphics[width=17cm]{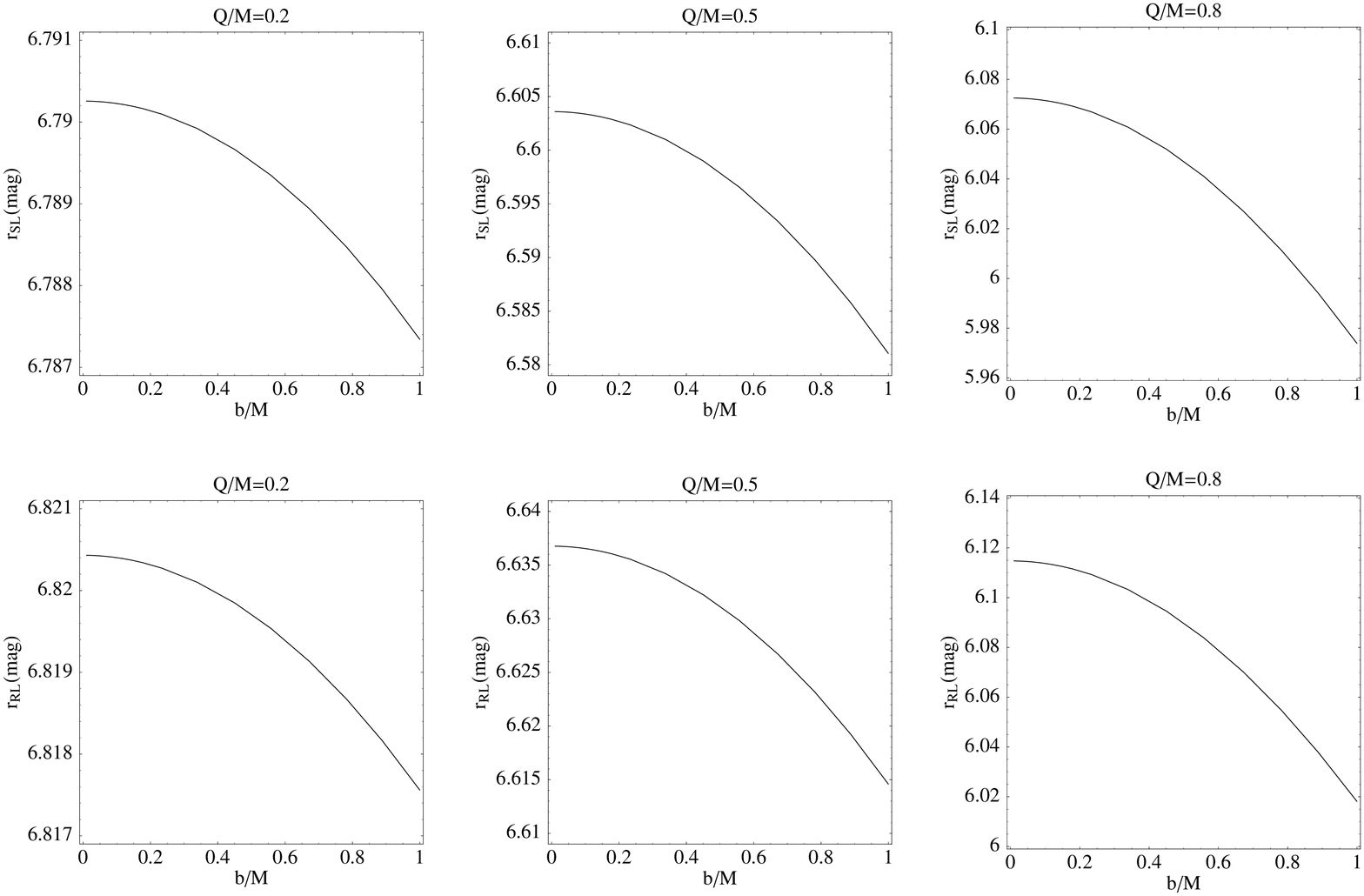}
\vspace{-1cm}
\end{center} 
\vspace{0cm} 
\caption{The values in magnitudes of $r_{\scriptscriptstyle{SL}}\mathrm{(mag)}=
2.5\log r_{\scriptscriptstyle{SL}}$ (upper panel) and 
$r_{\scriptscriptstyle{RL}}\mathrm{(mag)}=2.5\log r_{\scriptscriptstyle{RL}}$ 
(lower panel) are plotted as functions of the Born--Infeld parameter $b$ for 
different values of the charge $Q$. When $b=0$ the results corresponding to 
Reissner--Nordstr\"{o}m black holes are obtained. The values for Schwarzschild 
geometry are $r_{\scriptscriptstyle{SL}}^{Schw}\mathrm{(mag)}\approx 6.8212$ 
and $r_{\scriptscriptstyle{RL}}^{Schw}\mathrm{(mag)}\approx 6.8509$.}
\label{fig4}
\end{figure}
As a first example, the Galactic center black hole is considered as 
gravitational lens. The black hole mass is $M=3.6\times 10^{6}M_{\odot }$ 
\cite{eisenhauer} and its distance from the Earth is $7.6$ kpc 
\cite{eisenhauer}. Although its charge was not measured yet, large values of 
charge are not expected, but for completeness the values of charge are chosen 
to cover the theoretically possible range ($0 \leq |Q| \leq M$). The results 
corresponding to the observables defined above are shown in Figs. \ref{fig2} 
and \ref{fig4}. The limiting angular 
position $\theta _{\infty }$ is about $24\, \mu \mathrm{as}$ and the 
outermost image is about $0.03-0.06\, \mu \mathrm{as}$ from 
$\theta _{\infty }$ for standard lensing and $0.7-0.9\, \mu \mathrm{as}$ for 
retrolensing. For a given charge $Q$, the angle $\theta _{\infty }$ decreases 
with $b$, indicating that the images are closer to the optical axis than for 
Reissner--Nordstr\"{o}m geometry. The quantity $s$, instead, is an increasing 
function of $b$, for both standard lensing and retrolensing, so the images 
have a greater separation between them than in the Reissner--Nordstr\"{o}m 
case. The decreasing values of the observable $r$ with $b$, in both lensing 
configurations, indicates that the first relativistic image is less prominent 
with respect to the others than in Einstein--Maxwell gravity. It is 
important to remark that the differences between Einstein--Born--Infeld and 
Reissner--Nordstr\"{o}m geometries are very small and they grow by increasing 
the absolute value of the charge. The relativistic images are highly 
demagnified, so a bright source with high alignment and instruments with 
great sensitivity are required in order to observe them. Angular resolutions 
of less than $1\, \mu \mathrm{as}$ are also needed to separate the first image 
from the others. As it was shown by Eiroa and Torres \cite{eitor}, 
retrolensing images would be easier to detect than relativistic images in 
standard lensing situations. The results for the second example, 
a black hole with $M=7M_{\odot }$ placed at the galactic halo 
with $d_{ol}=4$ kpc, are shown in Figs. \ref{fig3} 
and \ref{fig4}. This black hole could have formed by gravitational collapse of 
a star with mass $M>10M_{\odot}$ \cite{punsly}. From Fig. \ref{fig3}, it can 
be seen that the values of $\theta _{\infty }$ and $s$ are five orders of 
magnitude smaller than those corresponding to the supermassive black hole. The 
behavior of the lensing observables with $b$ is similar to the other example 
analyzed above.

\section{Conclusions}

In this work, Einstein--Born--Infeld black holes were studied as gravitational 
lenses. In nonlinear electrodynamics, as a consequence of the self interaction 
of the electromagnetic field, photons follow null geodesics of an effective 
metric instead of those corresponding to the background geometry. The strong 
deflection limit coefficients $c_{1}$ and $c_{2}$ were numerically obtained 
from the effective metric, and they were subsequently used to find 
analytically the positions and magnifications of the relativistic images. 
The model was applied to the black hole in the Galactic center and to a 
low mass black hole at the Galactic halo. For a given 
value of charge, the innermost images are closer to the optical axis and the 
separations between the images grow as the Born--Infeld parameter $b$ 
increases. The images are highly demagnified and the outermost one is less 
prominent with respect to the others for increasing values of $b$. The 
differences with Reissner--Nordstr\"{o}m black hole lenses are 
very small and become more important for large values of charge. The 
gravitational lensing effects could be in principle used to discriminate 
between different black hole models, but the tiny differences between 
Einstein--Maxwell and Einstein--Born--Infeld spherically symmetric spacetimes 
will be extremely difficult to detect. The astronomical observation 
of these effects is beyond present technical capabilities and it will be 
a challenge for future facilities. Detailed discussions about the 
observational prospects of strong deflection lensing are given in 
Refs. \cite{bozman2,zakharov}.

\section*{Acknowledgments}

The author would like to thank Rafael Ferraro and Santiago E. Perez Bergliaffa 
for helpful discussions, and to an anonymous referee for useful comments that 
served to improve this article. This work has been partially supported by UBA 
(UBACYT X-103).

\end{document}